\documentclass[aps,prd,twocolumn]{revtex4}
\usepackage[per-mode = fraction]{siunitx}
\usepackage{graphicx}
\usepackage{amsmath}
\usepackage{amssymb}
\usepackage{color}

\usepackage[T2A]{fontenc}
\usepackage[utf8]{inputenc}
\usepackage[english,russian,ukrainian]{babel}

\begin{document}

\title{Entropy per particle spikes in the transition metal dichalcogenides}

\author{V.O. Shubnyi}
\affiliation{Department of Physics,
Taras Shevchenko National University of Kiev,
6 Academician Glushkov ave.,
Kiev 03680, Ukraine}

\author{V.P. Gusynin}
\affiliation{Bogolyubov Institute for Theoretical Physics, National Academy of Science of Ukraine, 14-b
        Metrolohichna Street, Kiev 03680, Ukraine}

\author{S.G. Sharapov}
\affiliation{Bogolyubov Institute for Theoretical Physics, National Academy of Science of Ukraine, 14-b
        Metrolohichna Street, Kiev 03680, Ukraine}

\author{A.A. Varlamov}
\affiliation{CNR-SPIN, University ``Tor Vergata'', Viale del Politecnico 1, I-00133 Rome, Italy}

%\ead{sharapov@bitp.kiev.ua}
%\email{sharapov@bitp.kiev.ua}
\vspace{10pt}

%\begin{indented}
%\item[]March 2015
%\end{indented}

\begin{abstract}
\selectlanguage{english}
We derive a general expression for the entropy per particle as a function of chemical potential, temperature and
gap magnitude for the single layer transition metal dichalcogenides. The electronic excitations in these materials can
be approximately regarded as two species of the massive or gapped Dirac fermions.
Inside the smaller gap there is a region with zero density of states where the dependence of the entropy per particle on the
chemical potential exhibits a huge dip-and-peak structure. The edge of the larger gap is accompanied by the discontinuity
of the density of states that results in the peak in the dependence of the entropy per particle on the chemical potential.
The specificity of the transition metal dichalcogenides makes possible the observation of these features at rather
high temperatures order of $\SI{100}{K}$. The influence of the uniaxial strain on the entropy per particle is discussed.

\iffalse
\selectlanguage{russian}
\vskip 0.5cm \par Мы изучаем

\selectlanguage{ukrainian}
\vskip 0.5cm \par Ми вивчаємо одношаровий
\fi
\end{abstract}

\selectlanguage{english}

%\pacs{73.22.−f, }
% 73.22.−f Electronic structure of nanoscale materials and related systems

%\keywords{Landau levels, graphene, strain}

\maketitle

\section{Introduction} \label{sec:intro}

We devote our work to the  memory of Alexei Alexeyevich Abrikosov. One of the topic of his research
at the end of the last millennium \cite{Abrikosov1998PRB} was the unusual magnetoresistance, linear in magnetic field
and positive, observed  in nonstoichiometric silver chalcogenides. His approach was
based on the assumption that these substances are gapless semiconductors with a linear energy spectrum
discovered by himself with coauthors in sixties \cite{Abrikosov1970JETP}.

This work of Abrikosov had drawn attention of two authors of the present work (VG and SS) to the various
realizations of the Dirac fermions in condensed matter systems. It was impossible to foresee that
the discovery of graphene in 2004 would make the Dirac fermions in condensed matter one of the hottest
topics of research for decades.

Another lesson that one may learn studying the scientific heritage of
Alexei Abrikosov is to focus on the theoretical results that are closely related to experiment.
He always taught that the article must be finished by the formula, which can be checked by experimentalist.
%Following his advice here we present a study of the entropy per particle,
%$s =\partial S/\partial n$, where $S$ is the entropy per unit volume and $n$ is the electron density,
%that was found wittily how to be measured experimentally by the authors of Ref.~\cite{Pudalov2015NatComm}.
Following his advice here we present a study of the entropy per particle $s =
\partial S/ \partial n$ ($S$ is the entropy per unit volume and $n$ is the
electron density) for which a witty approach for experimental measurement was
discovered by Kuntsevich \textit{et al.} \cite{Pudalov2015NatComm}.
In spite of the fundamental character of entropy  that  characterizes thermodynamics, heat transfer, thermoelectric properties
of  many-body systems, it is always hard to measure it directly.
%The recent experiment  \cite{Pudalov2015NatComm} is not an exception, because to be more exact
%the measured in a 2D electron gas quantity is the temperature derivative of the chemical potential,
%$\partial \mu /\partial T$.
The recent experiment \cite{Pudalov2015NatComm} is
not an exception as the quantity measured directly in a 2D electrong gas is the
temperature derivative of the chemical potential, $\partial \mu / \partial T$.
The key idea of the authors of experiment \cite{Pudalov2015NatComm}  is that
modulation of the sample temperature changes the chemical
potential and, hence, causes recharging of the gated structure,
where the 2D electrons and the gate act as two plates of a capacitor.
Therefore, $\partial \mu /\partial T$ is directly determined in the experiment from
the measured recharging current.
The Maxwell relation is then allows to equate both derivatives
\begin{equation}
\label{entropy-part}
s = \left( \frac{\partial S}{\partial  n} \right)_T = -\left( \frac{\partial \mu }{\partial T} \right)_n.
\end{equation}

It was theoretically predicted \cite{Varlamov2016PRB} that in
a quasi-two-dimensional electron gas (2DEG) with parabolic dispersion, the entropy per electron
exhibits quantized peaks when the chemical potential crosses the size quantized levels.
The amplitude of such peaks in the absence of scattering depends only on the
subband quantization number and is independent of material parameters,
shape of the confining potential, electron effective mass, and temperature.

Very recently we studied \cite{Tsaran2017SciRep} the behavior of $s$ as a function
of chemical potential, temperature and  gap magnitude for the gapped Dirac materials.
A special attention was paid to low-buckled Dirac materials \cite{Liu2011PRL,Liu2011PRB}, e.g.
silicene \cite{Kara2012SSR} and germanene \cite{Acun2015JPCM}.
The dispersion law in these materials writes
\begin{equation}
\label{silicene-spectrum}
\epsilon_{\eta s} (k) = \pm \sqrt{\hbar ^{2} v_{F}^{2} k^{2} + \Delta_{\eta \sigma} ^{2}},
\end{equation}
where $\eta = \pm 1$ and $\sigma = \pm 1$ are the valley and spin indices, respectively.
Here  $v_{F}$ is the Fermi velocity,
$k$ is the wavevector, and the valley- and spin-dependent gap,
$\Delta _{\eta \sigma }=\Delta _{z}-\eta \sigma \Delta _{\text{SO}}$, where $\Delta _{\text{SO}}$ is the
material dependent spin-orbit gap caused by a strong intrinsic spin-orbit interaction.
It can have a relatively large value, e.g. $\Delta _{\text{SO}} \approx \SI{4.2}{meV}$ in silicene and
$\Delta _{\text{SO}} \approx \SI{11.8}{meV}$ in germanene.
The adjustable part of the gap $\Delta_z = E_z d$, where $2 d$ is the separation between the two sublattices situated in different  planes,
can be tuned by applying an electric field $E_z$.
Accordingly, the density of states (DOS) reads
\begin{equation}
\label{DOS-silicene}
D\left( \epsilon\right) =f(\epsilon)\sum_{i=1}^N \theta \left(\epsilon^{2}-\Delta_i^{2}\right),
\end{equation}
where the function $f(\epsilon)$ is assumed to be a continuous even function of energy $\epsilon$
and
in the case of the discussed materials $N=2$ and $f(\epsilon) =  |\epsilon|/(\pi  \hbar ^{2} v_{F}^{2})$.
The DOS (\ref{DOS-silicene}) has  $4$ discontinuities at the points $\epsilon = \pm\Delta_i$, where
$i=1$ corresponds to $\eta = \sigma = \pm 1$ with $\Delta_1 = | \Delta _{\text{SO}} - \Delta _{z} |$ and
the second one with $i=2$ corresponds to $\eta = -\sigma = \pm 1$ with $\Delta_2 = |\Delta _{z}+ \Delta _{\text{SO}} |$.

One of the main results predicted in \cite{Tsaran2017SciRep} is that for $\mu = \pm \Delta_2$
($\Delta _{\text{SO}} ,\Delta _{z} >0$ was assumed) there is a peak of the height $s = \pm 2 \ln 2/3$
in entropy per particle when $T \to 0$.
The calculation of \cite{Tsaran2017SciRep}  shows that a peak at $\mu = \pm \Delta_2$
can still be seen for the temperature, $T \sim 10^{-2} \Delta_1$ for $\Delta_2 = 2 \Delta_1$.
Taking $\Delta_2 \sim \Delta _{\text{SO}}$, one estimates that the necessary temperature is the order of a few Kelvins.

Layered transition-metal dichalcogenides (TMDCs)  represent another class
of materials that can be shaped into monolayers, where similar effects might be observed.
Single layer TMDCs  with the composition MX$_2$
(where M = Mo, W is a transition metal, and X = S, Se, Te is a chalcogen atom)
are truly two-dimensional (2D) semiconductors with a large band gap of the order of \SIrange{1}{2}{eV} (see, e.g. Refs. \cite{Chhowalla2013NatChem,Kormanyos20152DMat}).
Consequently, one may expect that the peaks in entropy per particle can be seen at much higher temperatures.

The paper is organized as follows.  We begin by presenting in
section~\ref{sec:model} the model describing single layer TMDCs.
Since the full description of strained TMDCs is very complicated,
the effect of a uniform uniaxial strain is taken into account only via
scalar potential  spin-independent parts of the Hamiltonian.
In section~\ref{sec:entropy}  we discuss the DOS
and present an analytical expression for the entropy per particle
in TMDCs.  The results for the obtained behaviour of  the entropy per particle
are discussed in section~\ref{sec:results} and
conclusions are given in section~\ref{sec:concl}.

\section{Model}
\label{sec:model}

The low-energy excitations in monolayer TMDCs can be described by the following model Hamiltonian density
\cite{Capellutti2013PRB,Rostami2013Rostami,Liu2013PRB,Ridolfi2015JPCM,Kormanyos20152DMat}
\begin{equation}
\begin{split}
 H & = \sum_{\tau = \pm 1} H_\tau, \\
 H_\tau & = H_{D}^{\tau} + H_{2}
\end{split}
\end{equation}
where $\tau=\pm 1$ is the valley index,
$H_{D}^{\tau}$ is the linear in momentum in Dirac-like part \cite{Xiao20012PRL}
and $H_2$ is the quadratic part. The Dirac Hamiltonian contains free massive Dirac fermion, $H_{0}^\tau $,
and  spin-orbit term $H_{SO}^\tau$, $H_{D}^{\tau} = H_{0}^\tau + H_{SO}^\tau$.
The first term is
\begin{equation}
\label{H_0}
H_{0}^\tau = \hbar v_F(\tau k_x\sigma_x+k_y\sigma_y)+\frac{\Delta}{2}\sigma_z ,
\end{equation}
$\pmb{\sigma}$ are the Pauli matrices acting in the $2 \times 2$ ``band'' space,
$\sigma_0$ is the unit matrix, the Fermi velocity
$v_F = a t/\hbar \sim \SI{0.5d6}{m/s}$ with $t$ being the effective hopping integral
and $a$ is the lattice constant, the major band gap $\Delta \sim \text{\SIrange{1}{2}{eV}}$.
The inversion symmetry breaking results in the
the spin-orbit part of the Hamiltonian
\begin{equation}
\label{H_SO}
H_{SO}^\tau=\lambda_v\tau\frac{\sigma_0-\sigma_z}{2}s_z + \lambda_c\tau\frac{\sigma_0+\sigma_z}{2}s_z,
\end{equation}
where $s_z$ is the Pauli matrix for spin,
$2\lambda_v \sim \text{\SIrange{150}{500}{meV}}$ is the spin splitting at the valence band top
caused by the spin orbit coupling, $2 \lambda_c$ is the spin splitting  at
at the conduction band bottom. The DFT calculations \cite{Kormanyos20152DMat} show that
absolute value $2\lambda_v \gg  |2\lambda_c |\sim \text{\SIrange{3}{50}{meV}}$ and the sign
of $\lambda_c$ depends on the compound, $\lambda_c >0$  for MoX$_2$ and $\lambda_c < 0$ for WX$_2$ compounds.

The quadratic part of the Hamiltonian, $H_2$, contains the following diagonal terms
\begin{equation}
\label{H_2}
H_2 =\frac{\hbar^2 k^2}{4 m_e} (\alpha \sigma_0 + \beta \sigma_{z}),
\end{equation}
where $m_e$ is the free electron mass, and $\alpha \neq \beta$ are constants of the order of $1$.
Finally, as discussed in \cite{Capellutti2013PRB,Rostami2013Rostami,Liu2013PRB,Ridolfi2015JPCM,Kormanyos20152DMat}
more accurate approximations also include the trigonal warping terms.

The spin-up and spin-down components are completely decoupled,
thus the spin index $\sigma=\pm1$ is a good quantum number.
Neglecting the quadratic term (\ref{H_2}) we obtain
the dispersion laws for conduction and valence bands
\begin{equation}
\label{TMD-spectrum}
\begin{split}
\epsilon_{c,v}(k) & =  \frac{\lambda_v+\lambda_c}{2}\tau \sigma \\
& \pm\sqrt{ \hbar^2 v_F^2 k^2+ (\Delta-(\lambda_v-\lambda_c)\tau \sigma)^2/4}.
\end{split}
\end{equation}
This spectrum closely resembles  that of described by Eq.~(\ref{silicene-spectrum}) of massive  fermions
in low-buckled Dirac materials except to the first
valley- and spin-dependent term in Eq.~(\ref{TMD-spectrum}).
In the first approximation one can neglect the conduction band splitting and take $\lambda_c =0$ to arrive at the
simplest model \cite{Xiao20012PRL}, where  the conduction bands remain spin degenerate at $\mathbf{K}$ and $\mathbf{K}^\prime$
points and have small spin splitting quadratic in $\mathbf{k}$, whereas the valence bands are completely split,
\begin{equation}
\label{TMD-spectrum-simple}
\epsilon_{c,v}(k)=\frac{\lambda_v}{2}\tau  \sigma \pm\sqrt{\hbar^2 v_F^2 k^2+(\Delta-\lambda_v\tau \sigma)^2/4},
\end{equation}

Single layer TMDCs  can sustain deformations higher than
10\%  \cite{Bertolazzi2011ACSNano,Castellanos2012AdvMat}. The experimental possibility
to tune the band gap with strain has been proven for MoS$_2$ in
\cite{Conley2013NanoLett,Hui2013ACSNano,Castellanos-Gomez2013NanoLett,Zhu2013PRB}
and in WS$_2$ \cite{Wang2015NanoRes,Voiry2013NatMat,Georgiou2013NatNano}.
The full description of strained TMDCs is much more involved than that of graphene
and includes five different fictitious gauge fields as well as scalar potentials
entering spin-independent and spin-dependent parts
 of the Hamiltonian \cite{Rostami2015PRB}. Below we restrict ourselves by a qualitative estimate of the strain
effect on the properties of TMDCs and consider only the scalar potential term in the
spin-independent Hamiltonian (\ref{H_0}), viz.
\begin{equation}
\label{strain}
H_{\mathrm{str}} = \frac{D_{+}(\tensor{\varepsilon}) + D_{-} (\tensor{\varepsilon})}{2} \sigma_0 +
\frac{D_{+} (\tensor{\varepsilon}) - D_{-} (\tensor{\varepsilon})}{2} \sigma_3,
\end{equation}
where $\tensor{\varepsilon}$ is the strain tensor.
The explicit expressions for the diagonal terms $D_{\pm}$ are provided in \cite{Rostami2015PRB} and here
we only keep the linear in strain contributions neglecting the higher order terms
\begin{equation}
\label{D}
D_{\pm} = \alpha_2^{\pm} (\varepsilon_{xx} + \varepsilon_{yy}) ,
\end{equation}
with $\alpha_2^{+} = \SI{-3.07}{eV}$ and  $\alpha_2^{+} = \SI{-1.36}{eV}$.
The corresponding parameters %$\alpha_{2}^{s\pm}$
for the spin-dependent part are smaller by the three orders of magnitude, so that
the corresponding term can be safely neglected.
Assuming that the strain is a uniform uniaxial one, we can express $D_{\pm}$ via  $\varepsilon \equiv \varepsilon_{xx}$
($\varepsilon >0$ for tensile strain) and the
Poisson's ratio, $\nu$, \cite{Landau.book}  as follows $D_{\pm}= \alpha_2^{\pm} \varepsilon (1- \nu)$.
Thus in the present toy model the effect of strain is reduced to renormalization of
the chemical potential,
\begin{equation}
\label{mu-strain}
\mu \to \mu - \varepsilon (1-\nu)(\alpha_2^{+} + \alpha_2^{-})/2
\end{equation}
and the gap
\begin{equation}
\label{gap-strain}
\Delta \to \Delta + \varepsilon (1-\nu)(\alpha_2^{+} - \alpha_2^{-}).
\end{equation}
Setting $\nu =0$ one may estimate that 1\% tensile strain shifts $\mu$ by $\SI{22}{meV} $ and
$\Delta$ by $-\SI{17}{meV}$, respectively.

\section{Entropy per particle}
\label{sec:entropy}

As it was mentioned above, the entropy per particle is directly related to the
temperature derivative of the chemical potential at the fixed density $n$
(see Eq.~(\ref{entropy-part})). The latter can be obtained using the thermodynamic
identity
\begin{equation}
\label{derivative}
\left( \frac{\partial \mu}{\partial T} \right)_n = - \left( \frac{\partial n}{\partial T} \right)_\mu
\left( \frac{\partial n}{\partial \mu} \right)_T^{-1}.
\end{equation}
At thermal equilibrium, the total density of electrons is
\begin{equation}
\label{number}
n_{tot} (T,\mu)  =\int _{-\infty}^{\infty }d \epsilon D (\epsilon ) f_{\rm{FD}} \left(\frac{\epsilon -\mu }{T} \right),
\end{equation}
where $f_{\rm{FD}} ( x ) = 1/[\exp(x)+1 ]$ is the Fermi-Dirac distribution function and we set $k_B=1$.
Note that in the presence of the electron-hole symmetry
it is convenient to operate with the difference $n$ between the densities of electrons and holes instead of the total
density of electrons, as usually done for graphene \cite{Tsaran2017SciRep}.

One can show that in a close analogy with graphene and low-buckled Dirac materials
the DOS for TMDCs described by the approximate spectrum
(\ref{TMD-spectrum}) is
\begin{equation}
\label{TMD-DOS}
%\begin{split}
D(\epsilon)=\frac{1}{\pi(\hbar v_F)^2}\sum_{i = \pm 1}
\left| \epsilon- \epsilon_i \right|
\theta\left[(\epsilon- \epsilon_i)^2- \Delta_i^2\right].
%\end{split}
\end{equation}
Here we denoted
$\epsilon_i = i(\lambda_v + \lambda_c) /2 $ and
$\Delta_i = [\Delta-i(\lambda_v - \lambda_c)]/2$ with
$i=+1$ corresponding to $\tau = \sigma = \pm 1$ and
$i=-1$ corresponding to $\tau = -\sigma = \pm 1$.

Obviously  for $\lambda_c =0$  the resulting DOS corresponds to the spectrum
(\ref{TMD-spectrum-simple}). The DOS (\ref{TMD-DOS}) differs from the one described by the
equation (\ref{DOS-silicene}) by the presence of the energy shift, $\epsilon_i$, in the modulus
and in the argument of the $\theta$-function.
As a consequence the quantization of the entropy per particle, $s = \pm 2 \ln2/3$,
obtained in \cite{Tsaran2017SciRep}  for the low-buckled Dirac materials
does not occur in TMDCs.

\begin{figure}[!h]
\raggedleft{
\includegraphics[width=1.\linewidth]{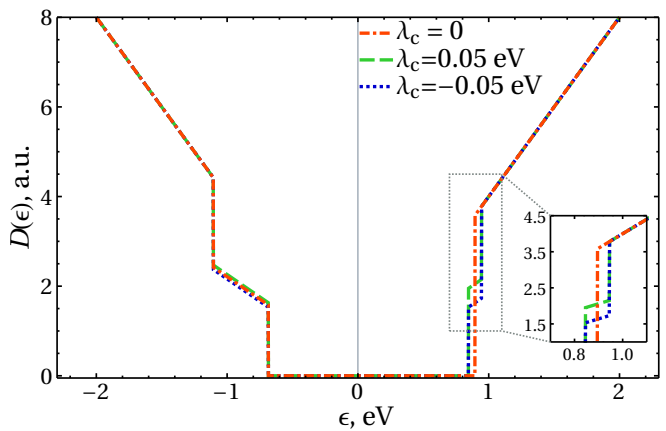}}
\caption{(Colour online) The DOS, $D (\epsilon)$, in arbitrary units versus energy in eV.
The parameters are $\Delta = \SI{1.79}{eV}$, $2\lambda_v= \SI{0.43}{eV}$. The dash-dotted (red) line $\lambda_c=0$,
long dashed (green) $\lambda_c= \SI{0.05}{eV}$, dotted (blue) $\lambda_c= \SI{-0.05}{eV}$.
} \label{fig:1}
\end{figure}

The behavior of the DOS given by Eq.~(\ref{TMD-DOS}) is illustrated in  Fig.~\ref{fig:1}.
To be specific, we took the values $\Delta = \SI{1.79}{eV}$, $2\lambda_v= \SI{0.43}{eV}$ corresponding to
the compound WS$_2$.
The constant $2\lambda_c$  for WS$_2$  is $\SI{-0.03}{eV}$ \cite{Kormanyos20152DMat}.
In order to demonstrate the role of this parameter we choose the larger value of $\lambda_c$.
Furthermore, we consider three possible cases: $\lambda_c =0$
is shown by  the dash-dotted (red) line, long dashed (green) line is for $\lambda_c = \SI{0.05}{eV}$,
dotted (blue) line is for  $\lambda_c = \SI{-0.05}{eV}$.
Note that in general {\it ab initio} density functional theory calculations \cite{Kormanyos20152DMat} predict that
$\lambda_c >0$ and  $\lambda_c <0$ correspond to MoX$_2$ and WX$_2$  compounds.
Going from the negative to positive energies we observe the first discontinuity of the DOS
at $\epsilon_{-1}^{-} = - \Delta/2 - \lambda_v = \SI{-1.11}{eV}$. It linearly goes down until  the second discontinuity
that occurs at  $\epsilon_{1}^{-} = - \Delta/2 + \lambda_v = \SI{-0.68}{eV}$. Their positions are independent of the value
of $\lambda_c$.
The DOS is zero inside the gap between $\epsilon_{1}^{-}$ and $\epsilon_{-1}^{+} =  \Delta/2 - \lambda_c$.
Then it increases linearly until the discontinuity at the energy,
$\epsilon_{1}^{+} =  \Delta/2 + \lambda_c$.
Obviously for $\lambda_c =0$ the last two discontinuities  become degenerate $\epsilon_{-1}^{+} = \epsilon_{1}^{+} = \SI{0.895}{eV}$.
For a finite $\lambda_c$  their ordering depends on the sign of $\lambda_c$.

The peculiarities of DOS  in  TMDCs beyond the Dirac approximations are discussed
in \cite{Scholz2013PRB,Iurov2017}. The quadratic part of the Hamiltonian  (\ref{H_2}) results in the
curving of the linear in energy pieces seen in Fig.~\ref{fig:1}.
Such curving is not essential and the  does not change the discontinuous character of the DOS function
that is responsible for the peaks in $s(\mu)$.

An advantage of the linearized approximation is that it resembles the case of gapped
graphene and allows to obtain rather simple analytical results. For example, one can derive the analytical expression
for the particle density (carrier imbalance)  \cite{Gorbar2002PRB} and find the derivative $\partial \mu/\partial T$
using Eq.~(\ref{derivative}). Its generalization  for the low-buckled Dirac materials was made in
\cite{Tsaran2017SciRep} (see also \cite{Iurov2017b}). The expression for the particle density in
TMDCs beyond the Dirac approximation is discussed in \cite{Iurov2017}, but it is not very practical for
obtaining the derivative $\partial \mu/\partial T$.

Differentiating Eq.~(\ref{number}) with respect to $T$ and $\mu$ and shifting the variable of integration
$\epsilon \to \epsilon + \epsilon_i$ for each term in the DOS (\ref{TMD-DOS})
 one obtains
\begin{equation}
\label{derivative-T-gen}
\left( \frac{\partial n_{tot}}{\partial T} \right)_\mu
= \int _{-\infty}^{\infty } \frac{d \epsilon  (\epsilon - \mu)  D (\epsilon )  }{4 T^2 \cosh^2 \frac{\epsilon - \mu}{2T}}
= \sum_{i = \pm 1} n_T (\mu_i,\Delta_i,T)
\end{equation}
and
\begin{equation}
\label{derivative-mu-gen}
\left( \frac{\partial n_{tot}}{\partial \mu }\right) _{T}
= \int _{-\infty}^{\infty } \frac{d \epsilon D (\epsilon )}{4 T \cosh^2 \frac{\epsilon - \mu}{2T}}
= \sum_{i = \pm 1}
n_\mu (\mu_i,\Delta_i,T),
\end{equation}
where $\mu_i=\mu-\epsilon_i$ is the shifted chemical potential. Since the corresponding integrands in
Eqs.~(\ref{derivative-T-gen}) and (\ref{derivative-mu-gen}) become formally the same as in the case of the low-buckled
Dirac materials \cite{Tsaran2017SciRep} we arrive at the final expressions
\begin{equation}
\label{derivative-T-2nd}
\begin{split}
n_T & (\mu,\Delta,T)  =  \frac{1}{\pi \hbar^2 v_F^2}  \left[
\frac{\Delta}{T} \frac{\mu \sinh (\Delta/T) + \Delta \sinh (\mu/T)}{\cosh (\Delta/T) + \cosh (\mu/T)} \right. \\
& + 2 T \mbox{Li}_2 \left(-e^{- \frac{\mu+ \Delta}{T}} \right) - 2 T \mbox{Li}_2 \left(-e^{ \frac{\mu- \Delta}{T}} \right)
+ \frac{2 \Delta \mu}{T} \\
& \left. - (\mu - 2 \Delta)  \ln \left( 2 \cosh \frac{\mu - \Delta}{2T} \right) \right. \\
& \left.  -
(\mu + 2 \Delta) \ln \left( 2 \cosh \frac{\mu + \Delta}{2T} \right)
\right].
\end{split}
\end{equation}
\iffalse
\begin{equation}
\label{derivative-T}
\begin{split}
& n_T (\mu,\Delta,T) =  \frac{1}{ \pi \hbar^2 v_F^2}
\left[ 2\Delta \ln \frac{1+\exp \left( \frac{\mu -\Delta }{T}%
\right) }{1+\exp \left( -\frac{\mu +\Delta }{T}\right) } \right. \\
& +2T\mbox{Li}_{2} \left( -e^{-\frac{\mu +\Delta }{T}}\right)
-2T\mbox{Li}_{2}\left( -e^{\frac{\mu -\Delta }{T}}\right) \\
& -\mu \ln \left( 2\cosh \frac{\mu -\Delta }{2T}\right) -\mu \ln \left( 2\cosh \frac{%
\mu +\Delta }{2T}\right)   \\
&+ \left. \frac{\Delta }{T}\frac{\mu \sinh (\Delta /T)+\Delta \sinh \mu /T}{%
\cosh \Delta /T+\cosh \mu /T}\right]
\end{split}
\end{equation}
\fi
and
\begin{equation}
\label{derivative-mu}
\begin{split}
& n_\mu (\mu,\Delta,T) =  \frac{1}{\pi \hbar
^{2}v_{F}^{2}}\left[ \frac{\Delta }{2}\left( \tanh \frac{\mu -\Delta }{2T}%
-\tanh \frac{\mu +\Delta }{2T}\right) \right.   \\
&+\left. T\left( \ln \left( 2\cosh \frac{\mu -\Delta }{2T}\right) +\ln
\left( 2\cosh \frac{\mu +\Delta }{2T}\right) \right) \right].
\end{split}
\end{equation}
$\mbox{Li}_2 (x)$ in Eq.~(\ref{derivative-T-2nd}) is the dilogarithm function.
As one can see, Eq.~(\ref{derivative-mu}) is symmetric with respect to the transformation $\mu \to -\mu$
or $\Delta\to-\Delta$. On the other hand  Eq.~(\ref{derivative-T-2nd}) is antisymmetric under change $\mu \to -\mu$ and
symmetric under $\Delta\to-\Delta$. The last property is checked using the identity for the dilogarithm function
\begin{equation}
{\rm Li}_2\left(-\frac{1}{z}\right)=-{\rm Li}_2(-z)-\frac{1}{2}\ln^2(z)-\frac{\pi^2}{6}.
\end{equation}

\section{Results}
\label{sec:results}

Basing on obtained Eqs.~(\ref{derivative}), (\ref{derivative-T-2nd}) and (\ref{derivative-mu}) one can investigate
the dependence $s(\mu)$ for the different cases.

\begin{figure}[!h]
\raggedleft{
\includegraphics[width=1.\linewidth]{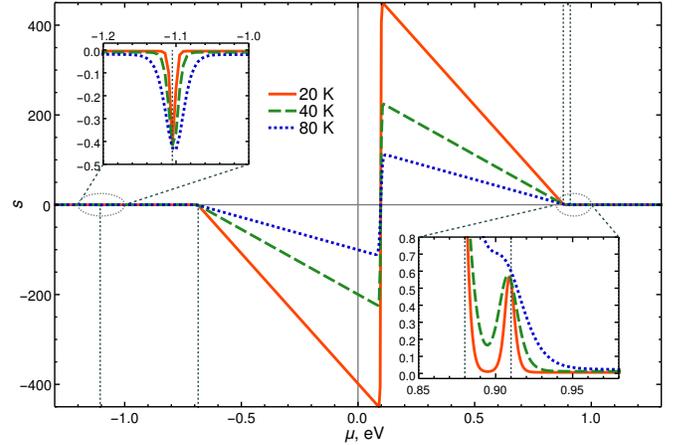}}
\caption{(Colour online) The entropy per electron $s$ vs the chemical potential $\mu$ in $\SI{}{eV}$ for three values of temperature.
The parameters are $\Delta = \SI{1.79}{eV}$, $2\lambda_v= \SI{0.43}{eV}$ and $\lambda_c= -\SI{0.015}{eV}$.
 }
\label{fig:2}
\end{figure}

Fig.~\ref{fig:2} is computed for the material parameters $\Delta$, $\lambda_v$ and $\lambda_c$
chosen for WS$_2$ compound. The dependence $s(\mu)$ is shown for three values of the temperature:
the solid (red) line is for $T = \SI{20}{K}$, the dashed (green) line is for $T = \SI{40}{K}$ and
the dotted (blue) line is for $T = \SI{80}{K}$. The vertical lines are
at the values of chemical potential
$\mu = \epsilon_{-1}^{-}, \epsilon_{1}^{-}, \epsilon_{1}^{+}, \epsilon_{-1}^{+}$
that correspond to the discontinuities of the DOS.
Comparing Fig.~\ref{fig:2} with the results presented in \cite{Tsaran2017SciRep}, one can
see that overall shape of $s(\mu)$ is similar for TMDCs and low-buckled Dirac materials,
although the details are different. For example, inside the gap for $\mu \in [\epsilon_1^{-}, \epsilon_1^{+}]$
the dependence of $s$ on the chemical potential exhibits a huge dip-and-peak structure in the
temperature vicinity of the point $\mu = (\lambda_v +\lambda_c)/2$. (The value $i=1$ corresponds to the smaller gap in
Eq.~(\ref{TMD-DOS})). This feature is
even more pronounced and sharp in TMDCs than in the other materials due to the larger ratio $\Delta/T$.
However in the low-buckled Dirac materials this structure was present in the temperature vicinity of the Dirac point,
$\mu =0$, because the whole dependence $s(\mu)$ was an antisymmetric function of $\mu$. This is obviously not the case of  TMDCs.
As discussed in \cite{Tsaran2017SciRep} the peak inside the gap is mainly due to the specific dependence
of the chemical potential on the electron density.

The presence of the second larger gap, $\Delta_2 > \Delta_1$, in silicene and similar materials
results in the emergence of the peak in $s(\mu)$ near the points $\mu = \pm \Delta_2$. Similarly
the discontinuities of the DOS given by Eq.~(\ref{TMD-DOS}) at $\mu = \epsilon_{-1}^{-},  \epsilon_{-1}^{+}$  associated with
a larger gap $i=2$ also result in the peaks in $s(\mu)$. They are shown in the inserts in Fig.~\ref{fig:2},
because they are much smaller in height. As explained above the value of $s$ at the peaks in the low temperature limit is not
equal to the quantized value $\pm 2 \ln 2/3$ expected for the low-buckled Dirac materials \cite{Tsaran2017SciRep}.
It is essential that both peaks can still be seen at rather high temperatures. The peak on the right starts to smear
at $T = \SI{80}{K}$, while the peak on the left can still be seen.

\begin{figure}[!h]
\raggedleft{
\includegraphics[width=1.\linewidth]{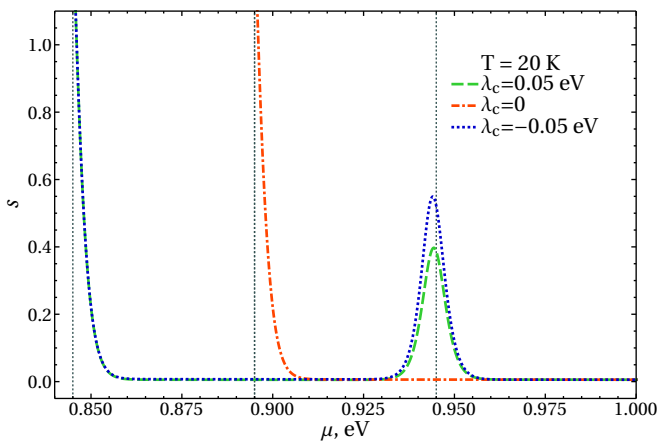}}
\caption{(Colour online) The entropy per electron $s$ vs the chemical potential $\mu$ in $\SI{}{eV}$ for three values of $\lambda_c = 0, \mp \SI{0.05}{eV}$.
The parameters are $\Delta = \SI{1.79}{eV}$, $2\lambda_v= \SI{0.43}{eV}$ and the temperature
$T =\SI{20}{K}$.
} \label{fig:3}
\end{figure}

It is shown in Fig.~\ref{fig:1} that for $\lambda_c =0$ the two discontinuities of the DOS merge at
$\mu = \epsilon_{-1}^{+} = \epsilon_{1}^{+}$. Then the positive peak in $s(\mu)$ disappears as can be seen on the
dash-dotted (red) in in Fig.~\ref{fig:3}.
As in Fig.~\ref{fig:2} the vertical lines correspond to the singularities of the DOS.
There is only one singularity for the dash-dotted (red) line at $\mu = \epsilon_{-1}^{+} = \epsilon_{1}^{+} = \SI{0.895}{eV}$.
For nonzero $\lambda_c$ there are two singularities shifted from this point to the left and right by $|\lambda_c| = \SI{0.05}{eV}$.
In this case the peak at the larger energy $\mu = \Delta/2 + |\lambda_c|$ is restored as can be seen on the
dotted (blue) line for $\lambda_c < 0$ and long dashed (green) line for  $\lambda_c >0 $.

Finally we consider how a uniform uniaxial strain would affect the results shown in Fig.~\ref{fig:2}.
We use Eqs.~(\ref{mu-strain}) and (\ref{gap-strain}) to model the dependence of chemical potential and gap $\Delta$
on the strain, respectively.
\begin{figure}[!h]
\raggedleft{
\includegraphics[width=1.\linewidth]{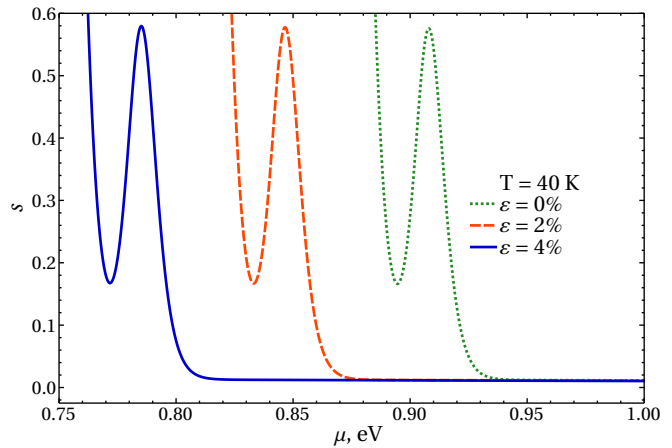}}
\caption{(Colour online) The entropy per electron $s$ vs the chemical potential $\mu$ in $\SI{}{eV}$ for three values of
strain. The parameters are
$\Delta = \SI{1.79}{eV}$, $2\lambda_v= \SI{0.43}{eV}$,  $\lambda_c= -\SI{0.015}{eV}$,
$\alpha_2^+ = \SI{-3.07}{eV}$,  $\alpha_2^- = \SI{-1.36}{eV}$ and the temperature
$T =\SI{40}{K}$. }
\label{fig:4}
\end{figure}
The dependence $s(\mu)$ is shown for three values of the strain:
the dotted (green) line is for $\varepsilon=0$,
the dashed (red) line is for $\varepsilon =2 \%$ and
the solid (blue) line is for $\varepsilon =4 \%$. As expected, the presence of strain results
in the movement of the peaks in $s(\mu)$.

\section{Conclusion} \label{sec:concl}

In the present work we had derived a general expression for the entropy per particle  as a function of the chemical potential,
temperature, and gap magnitude for the single layer transition metal dichalcogenides subjected to the  uniform uniaxial
strain. The spectrum of quasiparticle excitations of these materials is similar to that of the
low-buckled Dirac materials, viz. there is the valley- and spin-dependent gap
$\Delta_{\tau \sigma} = [\Delta- \tau \sigma(\lambda_v - \lambda_c)]/2$ in the spectrum.
The difference from the latter is that the whole spectrum is also shifted by a
valley- and spin- dependent constant $\epsilon_{\tau \sigma} = \tau \sigma (\lambda_v + \lambda_c) /2 $. This
introduces the hole-electron asymmetry in the band structure of TMDCs and makes the  resulting DOS (\ref{TMD-DOS})
asymmetric function of the energy.
When a small spin splitting at the conduction band bottom, $\lambda_c$, is taken into consideration
the DOS (\ref{TMD-DOS}) has  $4$ discontinuities: $2$ for the negative  and $2$ for the positive
energies. The positions of these discontinuities are not just at the energies $\pm |\Delta_{\tau \sigma}|$ with
$\tau = \sigma = \pm 1$ and $\tau = -\sigma = \pm 1$ due to the energy shift $\epsilon_{\tau \sigma}$.
It is demonstrated that inside the smaller gap there is a region with zero density of states where the dependence of
the entropy per particle on the chemical potential exhibits a huge dip-and-peak structure. The edge of the larger gap is
accompanied by the discontinuity of the density of states that results in the peak in the dependence
of s on the chemical potential. The specifics of the transition metal dichalcogenides makes the found  features
to be of the  ``high temperature'' nature, since they can be observed at rather high temperatures up to $\SI{100}{K}$.

Since the Seebeck coefficient is related to the temperature derivative of the chemical potential, the strong peaks in the
entropy per particle also indicate the same kind of singularities in the Seebeck coefficient in these materials. The latter
can be expected at the edge of the gaps and has the origin similar  to the electronic topological transitions
\cite{Varlamov1989AP,Blanter1994PR,Sharapov2012PRB}.

\begin{acknowledgments}

We thank A.O. Slobodeniuk for illuminating discussion.
We acknowledge the support of EC for the RISE Project CoExAN GA644076.
V.P.G. and S.G.Sh. acknowledge a partial support from the Program of Fundamental Research of the Physics
and Astronomy Division of the NAS of Ukraine No. 0117U00240.

\end{acknowledgments}

\end{document}